\documentclass{pasa}%

\usepackage{graphicx}

\usepackage{siunitx}

\title[GOTO with the LSST Science Pipelines: Forced Photometry]{Processing GOTO data with the Rubin Observatory LSST Science Pipelines II: Forced Photometry and light curves}

\author[Makrygianni et al.]{
L. Makrygianni,$^{1,2}$ J. Mullaney,$^{1}$ V. Dhillon,$^{1}$ S.
Littlefair,$^{1}$ K. Ackley,$^{3}$ M.J. Dyer,$^{1}$ J. Lyman,$^{4}$ K.
Ulaczyk,$^{4}$ R. Cutter,$^{4}$ Y.-L. Mong,$^{3}$ D. Steeghs,$^{4}$ D.
K. Galloway,$^{3,5}$ P. O'Brien,$^{6}$ G. Ramsay,$^{7}$ S.
Poshyachinda,$^{8}$ R. Kotak,$^{9}$ L. Nuttall,$^{10}$ E.
Pall\'e,$^{11}$ D. Pollacco,$^{4}$ E. Thrane,$^{3}$ S.
Aukkaravittayapun,$^{8}$ S. Awiphan,$^{8}$ R. Breton,$^{12}$ U.
Burhanudin,$^{1}$ P. Chote,$^{4}$ A. Chrimes,$^{4}$ E. Daw,$^{1}$ C.
Duffy,$^{7}$ R. Eyles-Ferris,$^{6}$ B. Gompertz,$^{4}$ T.
Heikkil\"a,$^{9}$ P. Irawati,$^{8}$ M. Kennedy,$^{12}$ T.
Killestein,$^{4}$ A. Levan,$^{4}$ T. Marsh,$^{4}$ D. Mata-
Sanchez,$^{12}$ S. Mattila,$^{9}$ J. Maund,$^{1}$ J. McCormac,$^{4}$
D. Mkrtichian,$^{8}$ E. Rol,$^{3}$ U. Sawangwit,$^{8}$ E.
Stanway,$^{4}$ R. Starling,$^{6}$ P. A. Str\o m,$^{4}$ S. Tooke,$^{6}$ K. Wiersema,$^{4}$
\affil{$^{1}$Department of Physics and Astronomy, University of Sheffield, Sheffield S3 7RH, UK}
\affil{$^{2}$The School of Physics and Astronomy, Tel Aviv University, Tel Aviv 69978, Israel}
\affil{$^{3}$School of Physics \& Astronomy, Monash University, Clayton VIC 3800, Australia}
\affil{$^{4}$Department of Physics, University of Warwick, Gibbet Hill Road, Coventry CV4 7AL, UK}
\affil{$^{5}$OzGrav: The ARC Centre of Excellence for Gravitational Wave Discovery, Clayton VIC 3800, Australia}
\affil{$^{6}$School of Physics \& Astronomy, University of Leicester, University Road, Leicester LE1 7RH, UK}
\affil{$^{7}$Armagh Observatory \& Planetarium, College Hill, Armagh, BT61 9DG}
\affil{$^{8}$National Astronomical Research Institute of Thailand, 260 Moo 4, T. Donkaew, A. Maerim, Chiangmai, 50180 Thailand}
\affil{$^{9}$Department of Physics \& Astronomy, University of Turku, Vesilinnantie 5, Turku, FI-20014, Finland}
\affil{$^{10}$University of Portsmouth, Portsmouth, PO1 3FX, UK}
\affil{$^{11}$Instituto de Astrof'{i}sica de Canarias, E-38205 La Laguna, Tenerife, Spain}
\affil{$^{12}$Jodrell Bank Centre for Astrophysics, Department of Physics and Astronomy, The University of Manchester, Manchester M13 9PL, UK}
}

\jid{PASA}
\doi{10.1017/pas.\the\year.xxx}
\jyear{\the\year}

\usepackage{aas_macros}
\usepackage{hyperref} 

\hypersetup{colorlinks,citecolor=blue,linkcolor=blue,urlcolor=blue}

\begin{document}

\begin{frontmatter}
\maketitle
\begin{abstract}
We have adapted the Vera C. Rubin Observatory Legacy Survey of Space and Time (LSST) Science Pipelines to process data from the Gravitational-Wave Optical Transient Observer (GOTO) prototype. In this paper, we describe how we used the Rubin Observatory LSST Science Pipelines to conduct forced photometry measurements on nightly GOTO data. By comparing the photometry measurements of sources taken on multiple nights, we find that the precision of our photometry is typically better than 20~mmag for sources brighter than 16 mag. We also compare our photometry measurements against colour-corrected PanSTARRS photometry, and find that the two agree to within 10~mmag (1$\sigma$) for bright (i.e., $\sim14^{\rm th}$~mag) sources to 200~mmag for faint (i.e., $\sim18^{\rm th}$~mag) sources. Additionally, we compare our results to those obtained by GOTO’s own in-house pipeline, {\sc gotophoto}, and obtain similar results. Based on repeatability measurements, we measure a $5\sigma$ L-band survey depth of between 19 and 20 magnitudes, depending on observing conditions. We assess, using repeated observations of non-varying standard SDSS stars, the accuracy of our uncertainties, which we find are typically overestimated by roughly a factor of two for bright sources (i.e., $<15^{\rm th}$~mag), but slightly underestimated (by roughly a factor of 1.25) for fainter sources ($>17^{\rm th}$~mag). Finally, we present lightcurves for a selection of variable sources, and compare them to those obtained with the Zwicky Transient Factory and GAIA. Despite the Rubin Observatory LSST Science Pipelines still undergoing active development, our results show that they are already delivering robust forced photometry measurements from GOTO data.
\end{abstract}

\begin{keywords}
Astronomical data analysis -- Astronomy software -- Surveys -- Photometry -- Light curves
\end{keywords}

\end{frontmatter}

%%%%%%%%%%%%%%%%%%%%%%%%%%%%%%%%%%%%%%%%%%%%%%%%%%

%%%%%%%%%%%%%%%%% BODY OF PAPER %%%%%%%%%%%%%%%%%%

\section{Introduction}

In the era of high-cadence, all-sky surveys, the processing and analysis of the vast amounts of resulting data is a major challenge. To date, a number of wide-field surveys have been commissioned to conduct large, repeated photometric surveys both in the optical (e.g., the Sloan Digital Sky Survey \linebreak (SDSS; \citealt{York2000}), the Palomar Transient Factory (PTF; \citealt{Law2009}, \citealt{Rau2009}) and Panoramic Survey Telescope and Rapid Response System (Pan-STARRS; \citealt{Hodapp2004}, \citealt{Chambers2016}) and the near-infrared (e.g., the Vista Variables in the Via Lactea survey conducted with the VISTA telescope; \citealt{Minniti10}). Repeated observations of the same parts of the sky enable time-domain studies, which are key for the identification and analysis of varying and transient astrophysical events. Along these lines, repeated surveys of large areas of sky are critical for statistical studies of varying and transient sources. Indeed, studies based on data from the aforementioned large surveys have shown that the properties that describe the level of variability of sources (such as variability timescales and amplitudes) correlate with other physical properties of the sources (e.g. \citealt{MacLeod2010}). Constraining such correlations will therefore enable us to use variability to infer other physical properties for large samples of astronomical sources. For example, measuring the variability of large numbers of AGN may, in the future, provide a further handle on the mass distribution of supermassive black holes (\citealt{Caplar2017},\citealt{Sanchez2018}). To achieve the full benefits of high-cadence, wide field surveys, it is critical that we are able to process large samples of photometric data efficiently and to a high level of precision.

In order to address the challenges presented by the data volume and rate delivered by wide-field, high cadence surveys, many of these projects have invested significant resources into the development of efficient data processing pipelines \linebreak (e.g., SDSS;\citealt{Lupton2001},Pan-STARRS; \citealt{Magnier2016}). The forthcoming Legacy Survey of Space and Time (LSST) to be conducted with the Vera C. Rubin Observatory (\citealt{Ivezic19}), represents a step-change in both data quantity and delivery rate. As a consequence, major efforts are currently being made by the LSST team to ensure that the data processing pipelines -- the LSST Science Pipelines (hereafter referred to as simply the ``LSST stack''; \citealt{Juric2017}) -- are capable of handling the data flow from the telescope.\footnote{The LSST stack software is available at https://github.com/lsst} The pipeline will deliver both of the main LSST data products i.e. those from the nightly processing and the annual releases.

Rather than being a single-purpose pipeline, however, the LSST stack has been designed to be adaptable for surveys conducted by facilities other than the Vera Rubin Observatory. Indeed, the Hyper Suprime-Cam Subaru Strategic Program (HSC-SSP;\citealt{Aihara2018a}), which is in the process of conducting a deep, multi-band imaging survey of selected fields, is also using a version of the LSST stack (\texttt{hscPipe}; \citealt{Bosch2018}) to process the data from that survey.

With this in mind, we have adapted the the LSST stack to process data from the Gravitational wave Optical Transient Observatory (GOTO; Steeghs et al. in prep.) -- a wide-field (currently $\sim$40 sq. deg), high-cadence survey telescope based on La Palma whose primary scientific objective is the identification of optical counterparts of gravitational wave events. The GOTO collaboration has developed their own in-house processing pipeline that has been optimised for the rapid follow-up of gravitational wave events. However, while GOTO's survey depth is shallower ($\sim20$ mag limit for a 3 minute exposure in dark time) than the LSST, both have a similar single-pointing field-of-views and cadences, meaning that the LSST stack is a viable alternative pipeline for non-primary science data products. In order to process GOTO data using the LSST stack, we have developed our own ``obs package'', \texttt{obs\_goto}, which is described in more detail in \cite{Mullaney20}.

Perhaps the most important data to come out of repeated, wide area surveys such as GOTO and the LSST is that of lightcurves; i.e., time-series data that describes how the flux of an astronomical object changes over time. In order to obtain lightcurves, flux measurements of the same object extracted from multiple observations must be associated with one another. One way of achieving this is via positional matching, in which a ``blind'' source detection (e.g., SExtractor; \citealt{BA1996}) algorithm is run on each incoming science exposure and common sources (i.e., those associated with the same physical object) are identified as those that lie within a given matching radius of each other in each observation. This method is, however, subject to a number of issues such non-detection in a survey because of low S/N in a given exposure (in which case, it is possible that a detected neighbouring object may be incorrectly matched, especially in a crowded field) and deblending failures. In an attempt to address these issues, the technique of ``forced photometry'' was developed. With this method, photometric measurements (e.g. flux) of sources are performed with the positions (and, if necessary, other parameters such as shape) fixed at those specified in a reference catalogue. Using this method we can, to a degree, mitigate the issues of non-detections or blended sources in a survey as the photometry will be measured for each position in the reference catalogue.

Motivated by the key role that forced photometry will play in the coming years with current and future multi-wavelength wide field surveys, we investigate the performance of the LSST stack's forced photometry task on wide-field survey data obtained by GOTO. Using this method we are also able to assess GOTO's photometric performance and compare against results obtained via ``blind source photometry'' i.e. photometry measurements of those sources identified in GOTO images via standard ``blind'' source detection, such as that performed by SExtractor (\citealt{BA1996}). Throughout this study, we use v18.01 \linebreak  (released July 2019) of the LSST stack, which was the most up-to-date version when we began processing our data.  It is important to note, however, that v18.01 of the LSST stack utilises the now near-obsolete “Generation 2” Butler to organise and retrieve data, which at the time of writing has largely been replaced by the ``Generation 3'' butler.

In this paper, we report on the forced photometry results we obtain by processing GOTO data using the LSST stack, with a particular emphasis on the quality of the photometry measurements. In the following section we provide a brief description of the GOTO survey, while in section 3 we give an overview of the forced photometry task and also how we filter bad data points from our light curves. In section 4, we present the results of various quality-assurance tests of the forced photometry measurements. Finally in section 5, we summarise our work and present our conclusions.

\section{The Gravitational-wave Optical Transient Observer}

The GOTO prototype is located on the summit of El Roque de los Muchachos (La Palma, Spain). It consists of an array of 40~cm-diameter astrographs (f/2.5) attached to the same mount, with each astrograph equipped with a 50M pixel detector with a field-of-view of roughly 5 sq. degrees (and a corresponding pixel scale of 1.24 arcsec). At the time of writing, GOTO consists of eight astrographs (hereafter, unit telescopes, or UTs) which is the full complement for a GOTO mount, although the data described in this work was obtained prior to the second set of four UTs being added i.e. during the GOTO prototype phase. A planned second dome located alongside the first will host an identical mount resulting in a total of 16 UTs and a total field-of-view of 80 sq. degrees, enabling repeat observations of the whole observable sky every few nights. A southern node is planned to be located in Australia which will provide full sky coverage for the GOTO survey. GOTO's default observing mode is the so-called ``survey mode'', in which the sky is repeatedly observed in a systematic way. This can be interrupted at any time to undertake prioritised observations to follow-up a transient event, such as a gravitational wave event (\citealt{Dyer2020}).

The four UTs used to obtain the data analysed in this study were aligned such that they deliver a contiguous field-of-view of roughly 20 sq. degrees per mount pointing. Each UT is equipped with a filter wheel consisting of standard Baader R, G, and B filters, plus a broad L-band filter which covers the optical passband between $\sim$400-700nm. The L-band is used as the primary filter for the survey as it maximises the amount of light reaching the detectors in a given exposure. 

Every night the GOTO Telescope Control System (G-TeCS; \citealt{Dyer2018}, \citealt{Dyer2020}) decides whether to open the dome given various criteria, including the local weather conditions. The pilot controls the hardware and will stop the operations if conditions are not appropriate and it will close the dome. Should the conditions improve/deteriorate during the night, the full robotic system will automatically resume/pause operations. The observations are also controlled by this system, with a scheduler deciding, in real-time, the optimal observations to conduct to achieve the primary science objective (i.e., the detection and identification of the optical counterparts of gravitational wave events and other transient sources). 

To identify the optical counterparts of gravitational wave events, GOTO repeatedly surveys the whole observable sky to ensure that up-to-date reference images exist to compare against incoming follow-up observations. This repeated survey provides the opportunity for science projects beyond identifying the optical counterparts of gravitational wave events. These include, for example, time-domain astrophysics and transient detections via forced photometry or image differencing on repeated observations of the sky.

Included in the LSST stack is a suite of software that is capable of conducting both image differencing and forced photometry. This includes the production of a set of reference images and catalogues (see \citealt{Mullaney20}) and, as we describe next, the tasks required to perform forced photometry based on the positions of sources in the aforementioned reference catalogue. The LSST stack also includes software capable of conducting image differencing, although this is beyond the scope of this study. 

\section{Data and Pipeline products}

\begin{figure}
	\includegraphics[width=\columnwidth]{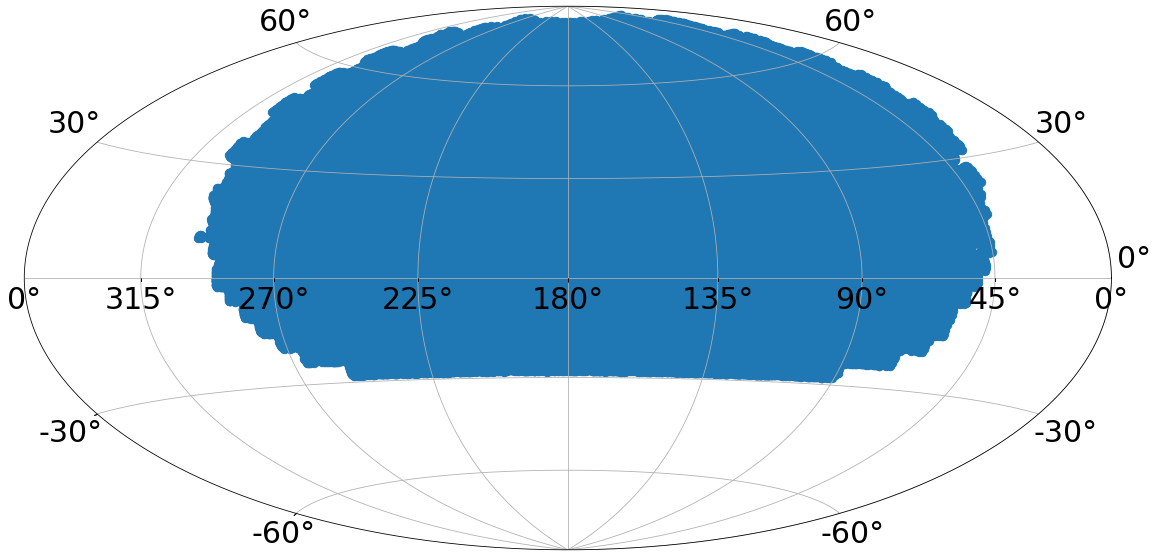}
    \caption{The region of the sky covered by the GOTO data that we processed using the LSST stack. These data cover the region spanning roughly -30 to 90 deg in declination and 15 to 315 deg in right ascension. This represents roughly 50\% of the sky observable at the location of the GOTO prototype on La Palma, Spain.}
    \label{fig:coverage}
\end{figure}

Typically, GOTO begins each night by obtaining a number of calibration frames; specifically bias, dark, and sky-flat frames. If the observing criteria are met (e.g., suitable weather conditions) then, after conducting a set of focusing exposures, it begins science observations. The data presented in this paper are the result of the nightly processing of raw images observed between the 24 February and the 25th October, 2019. Coadded images, from which reference catalogues are constructed, were produced by combining frames from dates spanning the 24 February to 12 March, 2019 (see \citealt{Mullaney20}). This selection resulted in reference images and catalogues spanning ${\rm 2~h \lesssim RA \lesssim 20~h}$ and ${\rm -20~deg \lesssim Dec. \lesssim 90~deg}$, which represents roughly 50\% of the sky observable from GOTO's location on La Palma and avoids the densest parts of the Galactic plane (see Fig. \ref{fig:coverage}). At each pointing, GOTO takes three back-to-back 60~s exposures which together are termed a ``visit'' . On coadding these three exposure we achieve an L-band $5\sigma$ magnitude limit of $\sim$20~mag (see section 4). While we describe the processing of GOTO images with the LSST software in detail in \cite{Mullaney20}, we feel it is important to highlight some of the key processing steps using the LSST stack here in order to provide some context.

First, the raw data are ingested into a database using the header information and master calibration frames that are produced by combining the individual bias, dark, and flat frames. This task can be done nightly or calibration frames from different nights can be combined. These master calibration frames are then used to correct science exposures for so-called instrument signatures, after which the individual science frames undergo background subtraction, PSF-characterisation, and astrometric and photometric calibration using a number of selected sources as astrometric and photometric standards. While the LSST stack can be fed bad pixel masks for flagging purposes, we did not provide these since they weren't available during GOTO's prototype phase when our data were taken. However, on visual inspection of the CCD images, it is clear that bad pixels represent far fewer than one in ten thousand pixels, and so we are confident that the impact of not including bad pixel masks has a minimal impact on our results. In this paper, we present the results from forced photometry using, as references, catalogues generated by running the LSST stack's detection algorithm on the coadded frames, and adopting a $5\sigma$ detection threshold (\citealt{Mullaney20}). In the following subsection we present the method we used to perform forced photometry on GOTO images using the LSST stack.

\subsection{Forced Photometry}
\label{frcphot}
Forced photometry is a technique that was developed to deal with issues arising from cross-matching between sky surveys conducted at different wavelengths and/or different telescopes. Depending on the shapes of their (observed-frame) spectral energy distributions, different sources may or may not be formally detected in different surveys. In such cases, a simple positional match may wrongly associate a source detected in one survey with a different, nearby, source in the other when, in reality, it is not formally detected in the latter. Further, sources that are close -- but still resolved as separate -- in one survey may be blended together in another survey, meaning that a simple positional match will associate all the resolved sources in one survey with the single unresolved source in the other. Forced photometry attempts to solve both these issues by performing photometry on one survey based on the positions of detected sources in another (e.g. \citealt{Bovy2012}, \citealt{Lang2014}, \citealt{Nyland2017}). In this respect, it is similar to the ``list-driven'' photometry technique described in \cite{Aigrain15}, although in our case we utilise our own catalogue derived from the coadded frames (see \citealt{Mullaney20}), rather than an external catalogue. Meaningful upper limits can be obtained via forced photometry in cases where a source would be formally undetected in a given survey. Of course, unless mitigating steps are taken, forced photometry can still suffer from flux contamination due to varying PSFs between different science frames. A further major drawback of forced photometry is that a transient source that does not exist in the reference catalogue would not be measured (unless, of course, it is associated with an existing, detected source, such as a supernova within a detected galaxy). To solve this problem, other techniques could be used, such as image differencing, which is beyond the scope of this study. Similarly, high proper motion stars may also be missed by forced photometry if the reference catalogue does not include proper motion information. As we are using our own catalogue derived from coadded GOTO frames as a reference, we do not have this information to hand. Since high proper motion stars represent an extremely small number of all astronomical sources, especially in the region outside the Galactic plane covered by our reference catalogue, we do not attempt to account for such sources.

To perform forced photometry on our incoming science frames, we use the LSST's {\tt forcedPhotCcd.py} task. This task finds the sources within the reference catalogue that overlap with the incoming science frames, and performs various (user-specified) photometric measurements at the positions of those sources. This approach means that every measurement in the incoming science frame is associated with an object ID within the reference catalogue. This association makes extracting light curves for a given object straightforward, as the user simply needs to specify the object ID of the source they are interested in.  

By default, {\tt forcedPhotCcd.py} performs forced photometry on every incoming science frame. However, GOTO takes multiple (usually three) back-to-back exposures for each pointing, which are grouped together according to their visit and CCD numbers. Each visit is identified via a unique identification number which associates it with a given pointing (see \citealt{Dyer2018} for further details). Since each exposure in a given visit is taken back-to-back, it is unlikely that there will be much change between exposures so, rather than perform forced photometry on every incoming exposure, we instead decided to coadd (using the LSST stack's {\tt snapCombine.py} task) the three back-to-back exposures to increase the depth of the forced photometry.\footnote{We choose to coadd, rather than take a median of, the three input exposures because taking a median can affect the PSF in non-trivial ways.} Prior to coaddition, however, each individual exposure requires instrument signature removal (ISR) and warping to a common WCS. We have therefore written our own wrapper for {\tt forcedPhotCcd.py} (named {\tt singleVisitDriver.py}) that processes and coadds individual frames prior to also performing forced photometry.    

We have made some further modifications relating to how the uncertainties associated with the photometric zero-point are calculated for each coadded exposure. In v18.01 of the LSST stack, the zero-point uncertainty is calculated as $\sqrt{\Sigma(1/\sigma^2)}$, where $\sigma$ is the so-called ``instrumental error'' associated with each measured source arising from photon noise. While this would hold true if the {\it only} source of error was counting statistics, in GOTO's case there are other sources of errors (e.g., varying conditions across the CCD) that would not get captured by this method. To obtain a more appropriate estimate of the uncertainty in the zero-point of a given frame($\sigma\textsubscript{zp}$), we instead use the standard deviation of the absolute difference between the instrumental magnitudes and the calibrated magnitudes (i.e., $\Delta m$) of the $N$ stars used to obtain the zero-point, i.e.,
\begin{equation}
\sigma\textsubscript{zp} = \sqrt \frac{{\Sigma \left(\Delta m - \overline{\Delta m}\right)^2}}{N}.
\end{equation}
Within the LSST stack, the error on the zero-point is then added in quadrature to the instrumental error for each measured source. A histogram of zero-point values for all our frames is presented in fig. \ref{fig:zeropoint}.

We have measured the total time taken to undertake the entire forced photometry task (i.e. from the calibration of the three back-to-back exposure to their coaddition to the forced measurements performed on the coadded frame) and report that it takes an average of $\sim$20~s per coadded frame.\footnote{ We used a Dual Intel Xeon E5-2697v3 2.60~GHz CPU with 28 cores/56 threads with access to 256 GB of RAM to process all our data. The reported time is the average wall-clock time to process a single coadded frame on a single core.} We note finally that, to speed up processing, we only perform forced photometry on the nightly frames (i.e., we do not also perform blind detection and measurement on the nightly frame). At the position of every reference source within the boundaries of a given coadded exposure we measure aperture photometry (using aperture radii: 5.58, 7.44, 11.16, 14.88, 29.76, and 59.52 arcsec which correspond to 4.5, 6, 9, 12, 24, and 48 pixels, respectively) and PSF photometry (using a PSF modelled using Principal Component Analysis; see \cite{Mullaney20} for more details). Our LSST stack-processed data is photometrically-calibrated using Pan-STARRS PS1 (\citealt{Magnier2016}) g-band PSF photometry, adopting appropriate colour terms to convert to the L-band (see \citealt{Mullaney20}). We use Pan-STARRS PSF photometry, as that is what is recommended for point sources, which the vast majority of calibration sources are. 

\subsection{Lightcurves}
\label{lightcurves}
The catalogues generated by {\tt forcedPhotDriver.py} contain the position of each reference source, aperture photometry measurements using various pre-defined aperture radii (see previous section) and their respective errors, and PSF photometry measurements and their errors. Forced photometry metadata are also generated containing information on the epoch, UT, filter, observed target or tile, visit number, zero-point and seeing of exposure. In this paper, we predominantly rely on aperture photometry for the results presented in section 4. We do, however, compare PSF photometry lightcurves against those extracted from aperture photometry as a way to estimate the quality of the PSF photometry. 

\begin{figure}
	\includegraphics[width=\columnwidth]{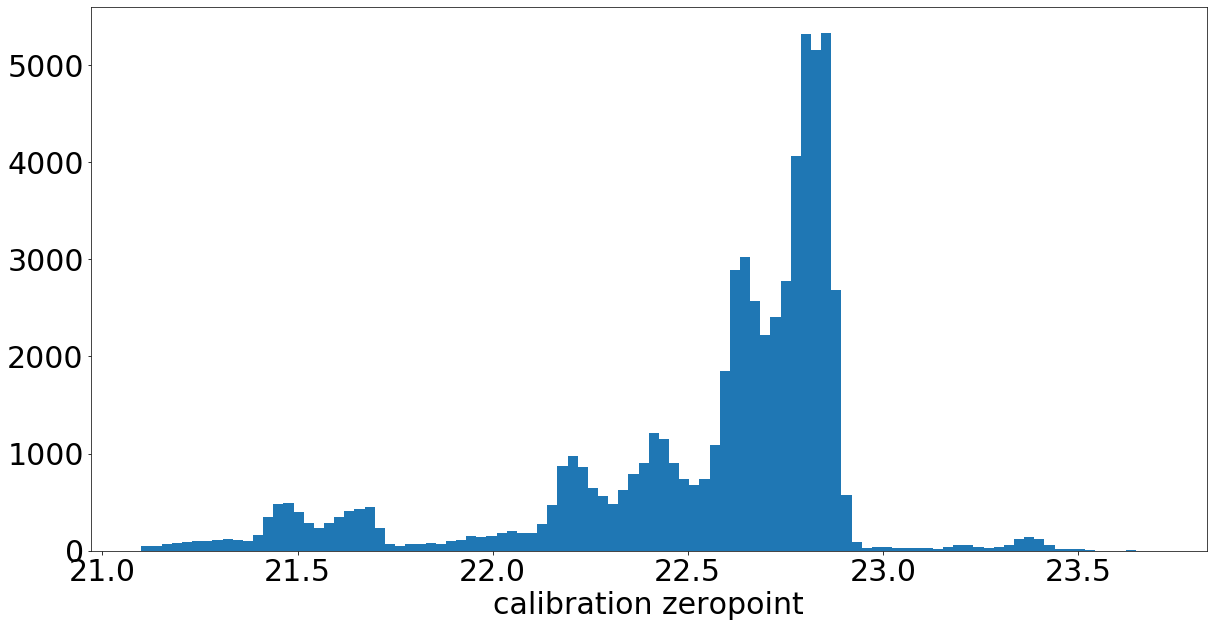}
    \caption{The distribution of photometric zero-points for the processed frames. The average and standard deviation of this distribution of 22.52 and 0.51. We exclude from further analysis any frames whose zero-points deviate from the average by more than three standard deviations. In further investigation, it was found that these were typically affected by poor observing conditions, including thin cloud cover.}
    \label{fig:zeropoint}
\end{figure}

Light curves are the main means by which the data from high cadence photometric surveys are analysed. As such, it is vitally important that the data used to construct lightcurves are reliable. There are various reasons, however, why this may not be the case. For example, data collected during nights of poor photometric quality, or instrumental or pipeline failures (which may or may not get flagged). It is therefore important to pre-process the light curves to ``clean'' the data of spurious photometric measurements which are not accounted-for by the reported uncertainties (e.g., in some cases, poor quality photometry is captured by the large uncertainties associated with the measured zero-point, but this is not always the case). In an attempt to remove poor quality data, we exclude any that arise from exposures with photometric zero-points that deviate by more than three standard deviations from the average. On further investigation these deviant zero-point values arose from frames that were affected by poor observing conditions such as thin cloud. The mean value of the photometric zero-point as measured by the LSST stack corrected for the exposure time for the data presented in this paper is 22.52 with standard deviation of 0.51 (see fig. \ref{fig:zeropoint}).

 Additionally, we exclude those sources that are flagged as interpolated, which, for the most part, arise due to the presence of saturated pixels in their footprint. In these cases the flux measurement fails because the masked pixels are not included when summing the flux within an aperture. If they were not excluded, such sources would have an underestimated measured flux which would not be reflected in the photometric uncertainty.
 
\section{Results}

In this section, we first compare the LSST software stack results for GOTO forced photometry against the blind source photometry from the Pan-STARRS DR1 catalogue as well as those measured by {\sc gotophoto} -- the in-house pipeline of the GOTO collaboration.\footnote{At the time of writing, {\sc gotophoto} does not conduct forced photometry.} We then assess whether the forced photometry results are self-consistent by testing the precision of the photometry across multiple nights. Using the photometric repeatability we obtain an estimate of the survey depth -- one that is independent of that estimated for the deeper reference images and catalogue described in \cite{Mullaney20}. In this section we also investigate the quality of our photometric uncertainties by assessing whether the reported uncertainties account for the scatter in the difference in measured photometry across multiple nights. Finally, we compare GOTO light curves generated by the LSST stack against those extracted from the Zwicky Transient Facility (ZTF; \citealt{Bellm2019}) and GAIA DR2 (\citealt{Clementini19}) databases for a number of known variable stars. 

\subsection{Photometry}
\label{photoqual}

\begin{figure*}
	\includegraphics[width=\columnwidth]{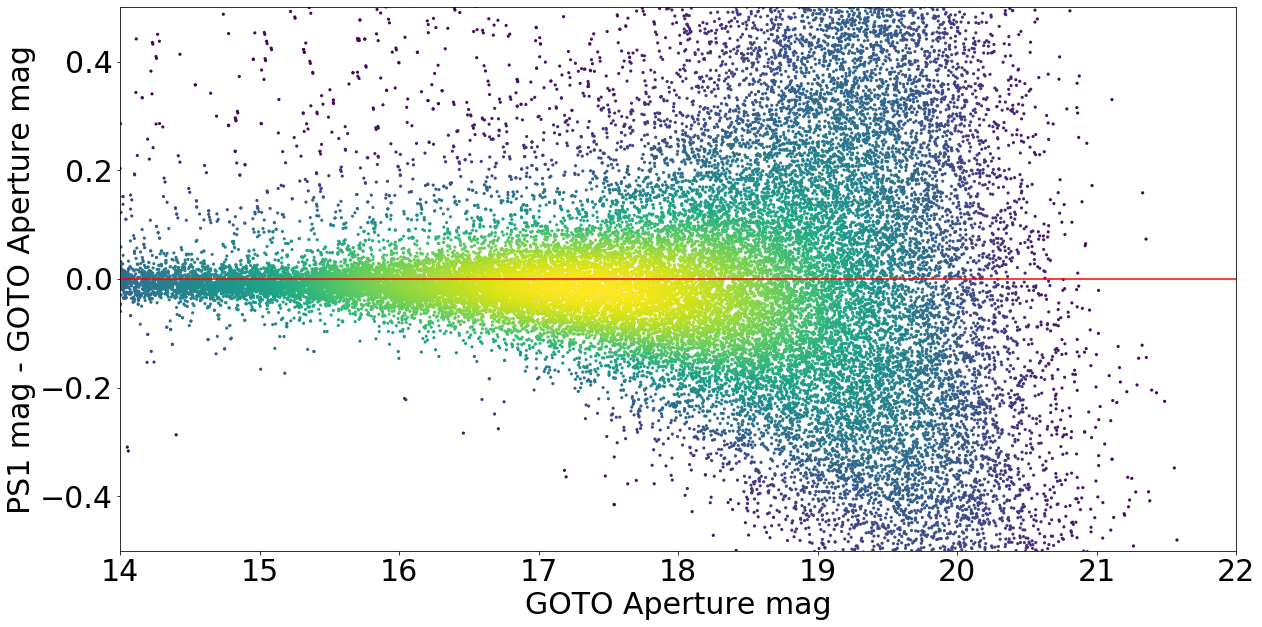}
	\includegraphics[width=\columnwidth]{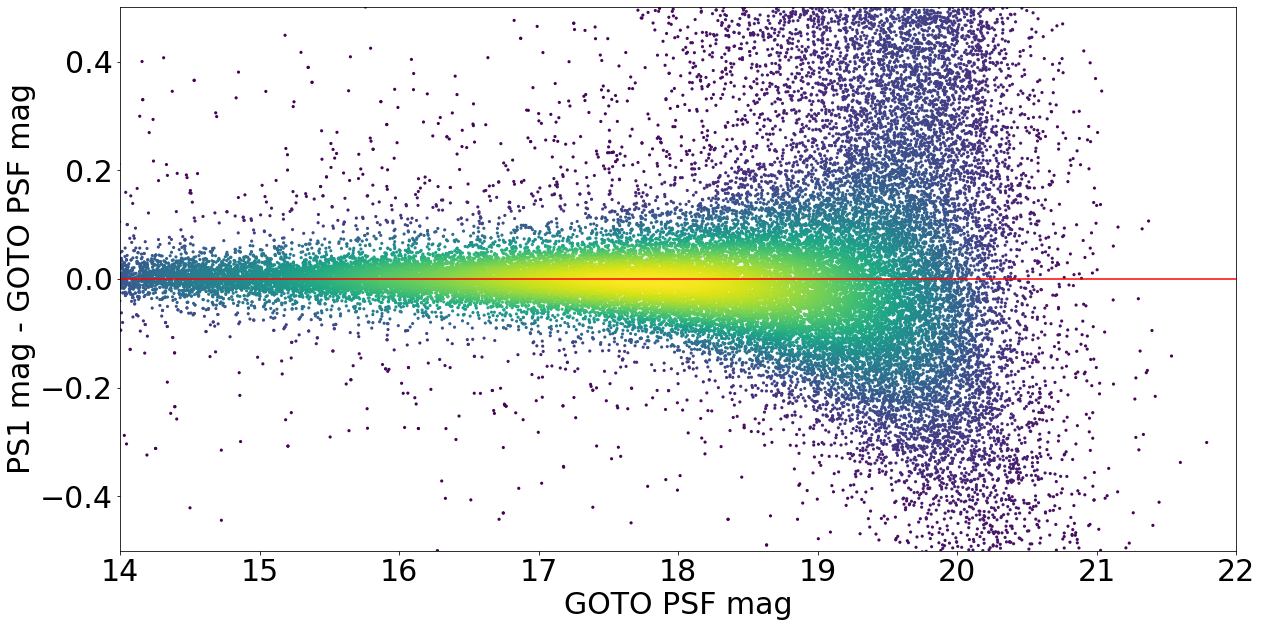}
	\centering
    \caption{The difference between GOTO photometry measured using the LSST stack and Pan-STARRS PSF photometry, plotted as a function of L-band magnitude. The left-hand plot shows the magnitude difference arising from GOTO aperture photometry, whereas the right-hand plot shows the difference arising from GOTO PSF photometry. Both plots show data arising from all four UTs for a single pointing.}
    \label{fig:LvsPS}
\end{figure*}

The first approach we take to estimate the quality of the photometry as measured by the LSST stack is to compare against the magnitudes reported in the Pan-STARRS DR1 catalog (\citealt{Magnier2016}). The Pan-STARRS photometry is calibrated using the \texttt{ubercal} method described in \cite{Schlafly2012}. Our choice of Pan-STARRS is motivated by the fact that it has a very similar sky coverage to the GOTO survey, but is significantly deeper than the GOTO survey, so all non-transient sources detected by GOTO {\it should} have a Pan-STARRS counterpart (the main exception being transient sources in GOTO). This means that we can obtain comparison statistics down to the detection limit of GOTO (i.e., we are not limited by the depth of Pan-STARRS). In Figure \ref{fig:LvsPS} we present plots showing the magnitude difference between the GOTO magnitudes measured by the LSST stack and the colour corrected Pan-STARRS g-band magnitude versus the GOTO magnitude. We present results from both aperture (11.16~arcsec; left) and PSF (right) photometry. This plot includes sources from all four CCDs for a single pointing, although we obtain similar results for all pointings (caveat those pointings filtered-out via the method outlined in \S\ref{lightcurves}).  This comparison with Pan-STARRS suggests that the PSF photometry is more precise for this particular epochal pointing, although as we shall see from the repeatability test, aperture photometry results are in general more accurate, and especially for sources brighter than $\sim17^{\rm th}$ magnitude. Both PSF and aperture photometry suggest that, for sources fainter than $18^{\rm th}$ magnitude down to the detection limit (i.e., $\sim$19.5), GOTO photometry as measured by the LSST stack is within 0.2 magnitude RMS of the Pan-STARRS photometry. Between 16th and 18th magnitude it is within $\sim$0.03-0.06 mag and for sources brighter the 16th magnitudes, the RMS is $\sim$0.01-0.02 mag.

\begin{figure*}
	\includegraphics[width=\columnwidth]{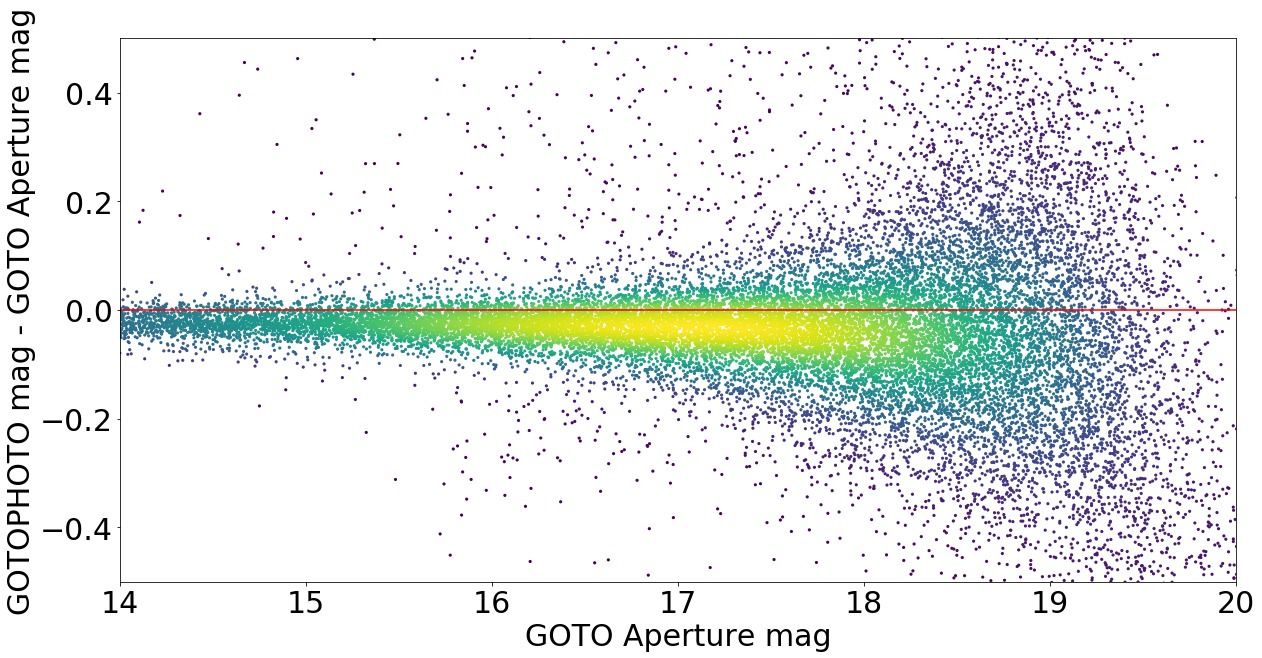}
	\includegraphics[width=\columnwidth]{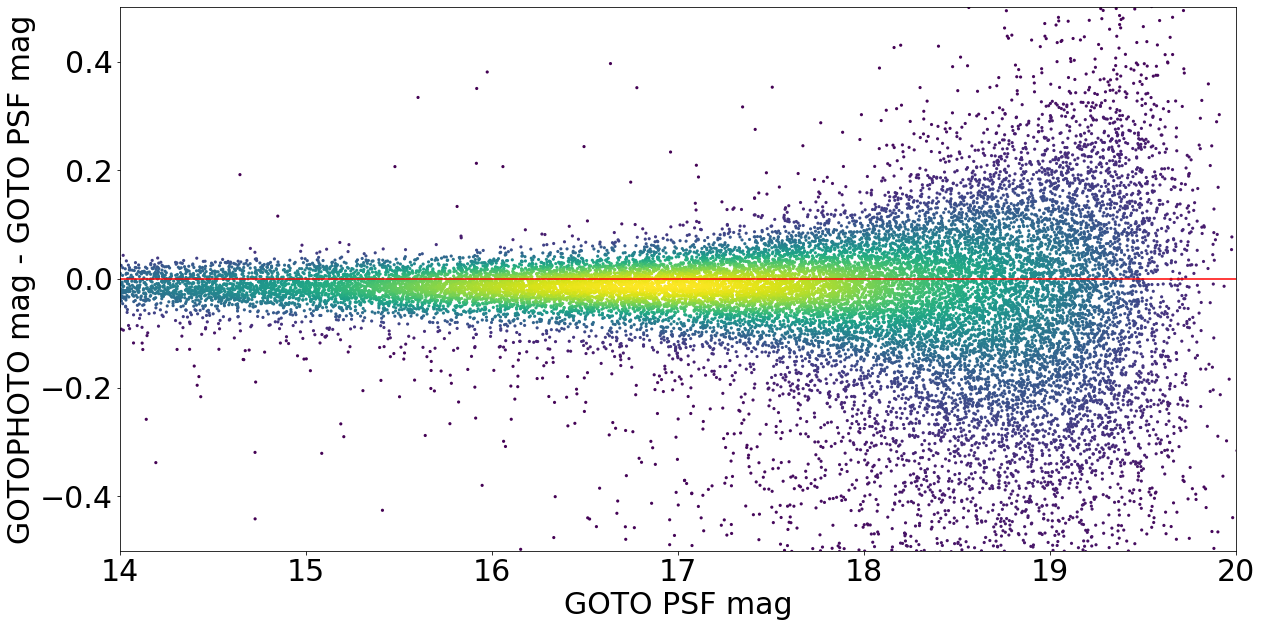}
	\centering
    \caption{As for fig. \protect\ref{fig:LvsPS}, but instead showing the difference between LSST stack-measured photometry, and that measured by  {\sc gotophoto}.}
    \label{fig:Lvsgotoflow}
\end{figure*}

We also compare outputs from the LSST stack against those obtained with {\sc gotophoto}, the in-house photometric pipeline developed by the GOTO collaboration.  {\sc gotophoto} uses SExtractor's (\citealt{BA1996}) {\tt MAG\_AUTO} photometry measurements for photometric calibration, which it compares against Pan-STARRS DR1 g-band PSF photometry to obtain photometric zero-points. Unlike the LSST stack-processed data, the version of {\sc gotophoto} used for our comparison does not apply colour terms to convert between Pan-STARRS g-band photometry measurements to GOTO's L-band (although there are some colour cuts on the stars chosen to calibrate and image to remove strong outliers, i.e., -0.5 < g-r < 1.0).\footnote{Colour terms will be implemented in future versions of {\sc gotophoto}} We note that here we are comparing forced photometry measurements (from the LSST stack) against measurements of sources obtained via blind detection (from {\sc gotophoto}). This caveat shouldn't be a concern for isolated sources, which form the vast majority, although it could mean that some sources are deblended in the blind catalogue, but not in the forced photometry catalogue (or vice versa). {\sc gotophoto}'s measurements are based on SExtractor's (\citealt{BA1996}) {\tt MAG\_AUTO} aperture photometry performed on the same (but median-combined, rather than coadded) back-to-back exposures as those we performed forced photometry on. In Figure \ref{fig:Lvsgotoflow} we present plots showing the magnitude difference between the GOTO magnitudes measured by the LSST stack and those measured by {\sc gotophoto} versus GOTO magnitude. From this comparison we note that the vast majority are within an RMS of 0.1~mag, even in the case of the faintest sources. There is, however, a systematic offset between the results of the two pipeline which is likely due to the application of colour terms when we process the data with the LSST stack (which are not applied to {\sc gotophoto} magnitudes). The PSF photometry appears once again to result in a smaller scatter than aperture photometry for this particular pointing, although we note that other pointings produce similar overall results. The accuracy of PSF photometry vs. aperture photometry is investigated using the photometric repeatability, which we consider next.

\begin{figure}
	\includegraphics[width=\columnwidth]{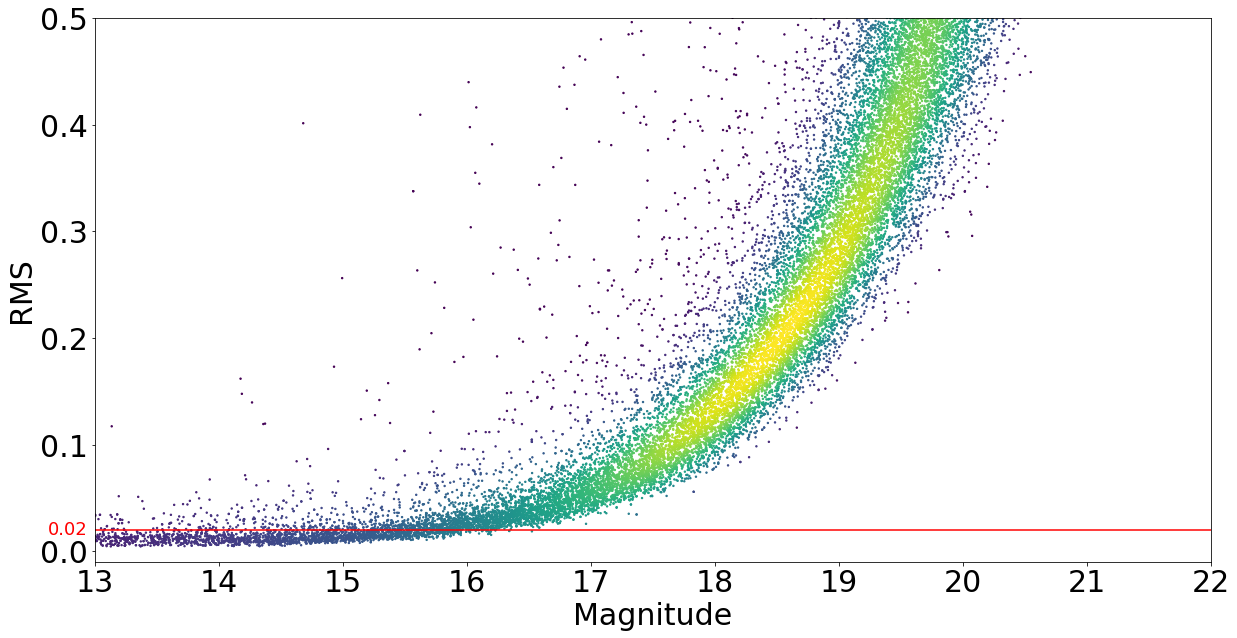}
	\centering
    \caption{Photometric repeatability for 11.16~arcsec aperture photometry as measured with the LSST stack, plotted as a function of L-Band magnitude. Each point represents a single reference source within a 4 UT pointing. We use the RMS of the photometry of these sources measured across multiple nights as our measure of repeatability; see Section \ref{photoqual} for details. A photometric precision of 0.02 mag (shown as the red line) is achieved for bright (i.e., $m_L\lesssim15$) sources. }
    \label{fig:repeatability}
\end{figure}

\begin{figure}
	\includegraphics[width=\columnwidth]{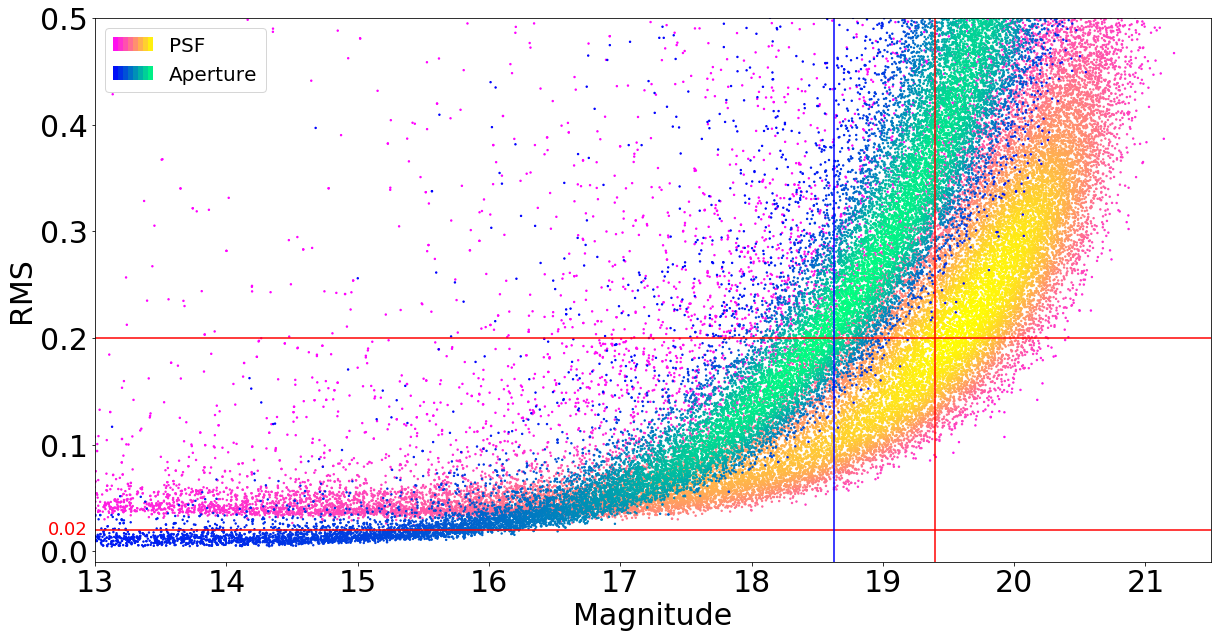}
	\centering
    \caption{As for fig. \protect\ref{fig:repeatability}, but now also showing the repeatability of PSF photometry for comparison. While PSF photometry is less precise for the brighter sources, it may be a better choice for fainter sources. We also use the inter-night RMS to estimate the survey depth by using 0.2 mags RMS as an estimate of the 5$\sigma$ detection threshold. This is shown as the horizontal red line at $RMS=0.2$~mag, which corresponds to L-band survey depths of 18.6 and 19.4 for 11.16 arcsec aperture photometry and PSF photometry, respectively.}
    \label{fig:psfvsAp}
\end{figure}

With photometric repeatability tests, we are investigating the level of consistency between multiple photometric measurements of the same source across multiple nights. We estimate this quantity for both types of photometry measurements obtained using our implementation of the LSST stack, i.e., 11.16 arcsec aperture and PSF photometry. We investigate how the photometric repeatability changes as a function of magnitude. To do this, we use data obtained from observations of a region of sky that has been visited the most number of times by GOTO between 2019-03-01 and 2019-07-31 (inclusive) and calculate the RMS of the magnitude of all sources in one pointing (from all four UTs). To clean the photometric data of spurious measurements we use a similar method using pixel flags as that outlined in section 3.2. Further, we remove measurements from any frame whose zero-point deviates by more than three standard deviations from 22.52 (i.e., the mean zero-point reported in \S\ref{lightcurves}). Both of these cleaning steps are straightforward to do using the metadata provided by the LSST stack. We note that we do not filter for known variable sources. Such sources will increase the measured inter-night RMS, but they represent such a small proportion of sources that we do not expect them to have any measurable effect on our repeatability measurements. 

In figure~\ref{fig:repeatability} we plot the inter-night RMS of aperture photometry measurements from the aforementioned observations. From this plot we see that for sources brighter than $15^{\rm th}$~mag the internal photometric precision, as measured by photometric repeatability, is typically below $\sim$0.02 mag (red line on the plot). The RMS increases with decreasing brightness due to the increase in size of the photometric uncertainties.
%RLCS: could refer to the red horizontal line in the figure at 0.02 here-DONE

%\begin{figure}
%	\includegraphics[width=\columnwidth]{figures/rms_phot}
  %  \caption{\bf DO WE NEED THIS PLOT? IT LOOKS JUST LIKE FIG. 6, BUT WITH LINES DRAWN ON. CAN'T WE JUST ADD LINES TO FIG. 6?}
    %\label{fig:rmsphot}
%\end{figure}

We compare the repeatability of PSF and aperture photometry in fig. \ref{fig:psfvsAp}. This plot shows that PSF photometry is less consistent between nights than aperture photometry for sources brighter than around $m_{\rm L} = 17$. This limitation is expected due to the various difficulties associated with performing PSF photometry on bright sources arising from, e.g., bright spikes or saturated pixels. PSF photometry, however, is found to perform better than aperture photometry for fainter sources. In the current GOTO prototype system, the PSF can vary over the field-of-view, especially at the edges. However, as shown in Fig. 5 of \cite{Mullaney20}, the LSST stack's PSF modelling software is able to account for this variation and, in general, does a good job of reproducing the PSF across the frame. 

In the aforementioned analysis we exclusively used 11.16 arcsec aperture, i.e. $\sim$2.5 times the typical FWHM of the PSF size ($\sim$4.5~arcsec), for aperture photometry. However, as mentioned in \S\ref{frcphot}, our implementation of the LSST stack returns measurements obtained with multiple different sized apertures. In general, we find that smaller apertures reproduce the inter-night RMS of PSF photometry more faithfully than larger apertures (i.e., smaller apertures result in a larger systematic RMS at brighter magnitudes, but smaller RMS values at fainter magnitudes). Considering this, it may be beneficial in the future to attempt to adjust the aperture to match the size of the PSF, at least when measuring faint sources.

\subsection{Survey depth from repeated photometry}

In \cite{Mullaney20} we characterised the depth and the detection completeness of the coadded images based on the average magnitude of a $5\sigma$ detected sources. This was further verified using injected sources. However, another way to obtain an estimate of the magnitude limit for the GOTO survey is from photometric repeatability. The 5$\sigma$ detection corresponds to a S/N of 5 and so to a flux RMS of $\sim$ 0.2 mags (\citealt{Masci2019}; note that in this regime we are dominated by random, as opposed to systematic, errors). Using aperture photometry measurements from the same set of observations as used in \S\ref{photoqual} we calculate the median magnitude of sources with a S\/N of between 4.5 and 5.5. As shown in Figure \ref{fig:psfvsAp}, this corresponds to a magnitude limit for the L-band of 18.6~mag. This value is brighter than the L-band magnitude limit of 19.6~mag for the coadded references exposures presented in \cite{Mullaney20}. This discrepancy arises from the fact that the analysis described in this paper is based on forced photometry performed on single visit frames. While each of these visit frames is produced by mean-combining three back-to-back exposures, each coadd (or part thereof) may have been produced from the combination of more than three exposures. 

One important issue relating to the estimation of the survey depth relates to the choice of the type of photometry used (i.e., aperture vs. PSF) and, in the case of aperture photometry, the size for aperture. This choice is especially relevant for the faint sources that we use to define the survey depth. In our case the 11.16~arcsec aperture will include more background flux than a smaller aperture, possibly leading to an overestimation of the brightness of the faint sources. We can see from Figure \ref{fig:repeatability} that if we use the PSF photometry for the fainter sources then we would obtain a limiting magnitude of 19.4~mag (note that PSF photometry was not performed by \citealt{Mullaney20}, so a comparison between forced and reference PSF photometry cannot be made).

\subsection{Observational photometric uncertainties}

\begin{figure}
	\includegraphics[width=\columnwidth]{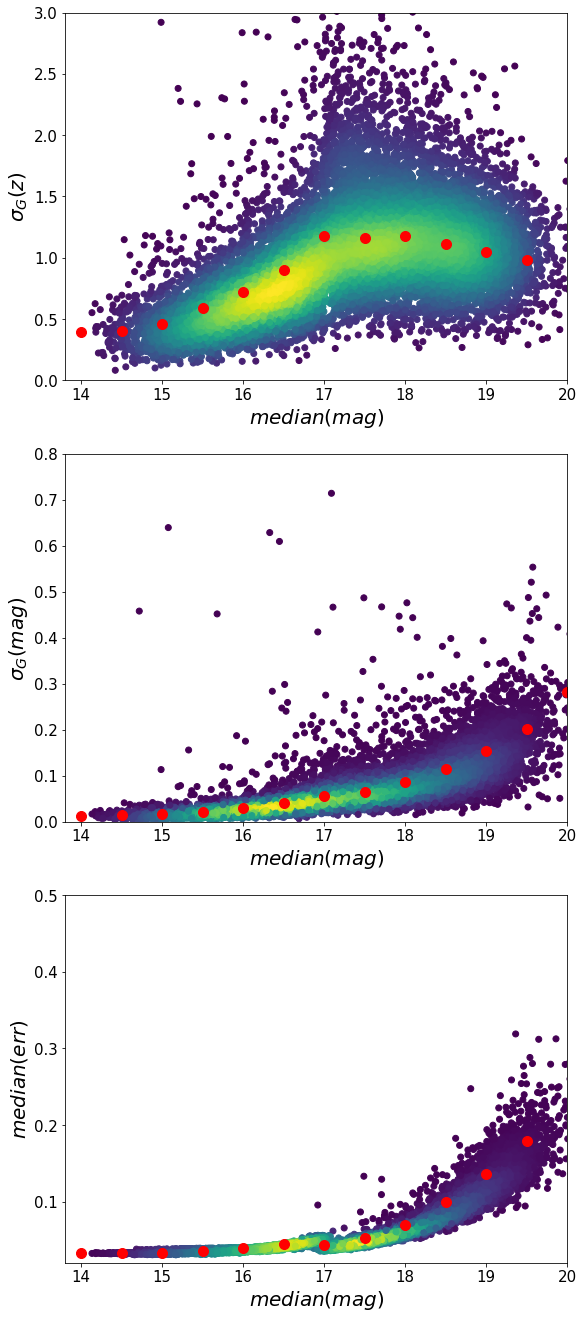}
    \caption{Plots used to assess the quality of the uncertainties of our photometry measurements. The top panel shows the robust standard deviation of the quantity of Eqn. \ref{zij} as a function of the median magnitude for the GOTO light curves from forced photometry. The middle shows the intrinsic scatter of each light curve measured as the robust standard deviation of the magnitude as a function of the magnitude and the bottom panel shows the median error of each light curve as a function of the magnitude. The method evaluates the photometric uncertainties following the method outlined in \protect\cite{Suberlak2017}.}
    \label{fig:errorplots}
\end{figure}

As well as assessing the quality of the absolute photometric measurements, it is important that we also assess their uncertainties. This quantity is particularly relevant for the analysis of variable sources, since we need to know whether differences in the measured photometry over multiple nights are physical in origin (i.e., genuine), or simply due to the statistical variances in our measurements. As such, we must carefully evaluate whether we are over or under-estimating our uncertainties.\footnote{An example of the effects of poor error estimation is presented in \citealt{Suberlak2017}, in which they find that the quasar variability levels observed in Catalina Real-time Transient Survey (CRTS) data (\citealt{Graham2014}) actually arises from underestimated errors.}

\begin{figure*}
	\includegraphics[width=\linewidth]{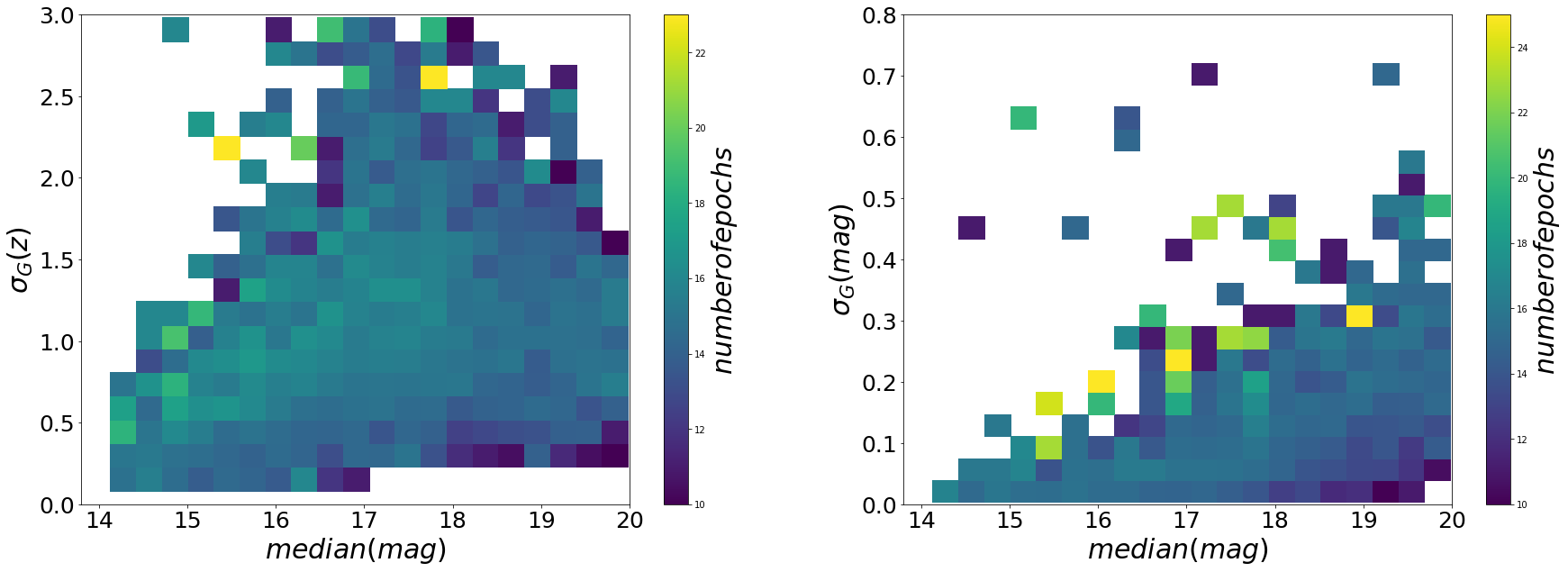}
    \caption{Standard deviation of the photometric scatter normalised for the reported photometric uncertainties (left) and intrinsic standard deviation (right) as function of magnitude and coloured by the number of epochs in the light curve. There is no obvious correlation for either of them with the number of epochs.}
    \label{fig:errorepochs}
\end{figure*}

As described in \S\ref{frcphot}, the uncertainties reported by the LSST stack on each photometric measurement are obtained by combining, in quadrature, instrumental uncertainties (i.e., those principally arising from photon noise) with calibration uncertainties (i.e., those arising from uncertainties in the zero-point). To assess the quality of these uncertainties we use the method outlined in \cite{Suberlak2017} that they use to assess the quality of the uncertainties reported by the CRTS. We repeatedly measure the  photometry of stars extracted from the \cite{Ivezic2007} catalogue of standard stars.\footnote{To create this catalogue, \cite{Ivezic2007} used repeat measurements of $\sim$1 million $m_r=14-22$ stars in SDSS Stripe82 to verify that they are non-variable.} For each standard star, $j$, covered by our repeat GOTO observations, $i$, we calculate $z_{ij}$:

\begin{equation}
 z_{ij} = \frac{m_{ij}-\overline{m_{ij}}}{\epsilon_{ij}}
\label{zij}
\end{equation} 
  
\noindent where $m_{ij}$ and $\epsilon_{ij}$ are the measured L-band photometries and associated uncertainties of source $j$ from observation $i$, and $\overline{m_{ij}}$ is the mean of all $m_{ij}$, averaged over $i$, weighted according to inverse uncertainty. We then take the standard deviation of these $z_{ij}$ values for each star, using the definition of standard deviation used in \cite{Suberlak2017}, which is less affected by outliers:

\begin{equation}
	\label{sdev}
\sigma_j(z) = 0.741\times IQR(z_{ij})
\end{equation} 

\noindent where $IQR(z_{ij})$ is the 25\%-75\% interquartile range of the $z_{ij}$ values over all observations, $i$, of each standard star, $j$. As such, there are $j$ $\sigma_j(z)$ values, i.e., one per star.

There are $\sim$13000 standard SDSS stars within the current GOTO reference catalogue used as the basis for our forced photometry. The r-band magnitude range of these stars is 14-20. In the top panel of Figure \ref{fig:errorplots} we plot $\sigma_j$ for each of these stars. For non-variable stars, it is expected that the $\sigma_j$ values would follow a distribution centred at unity and display no dependence on magnitude. However, as for CRTS, we find that this is not the case for GOTO photometry as measured by the LSST stack. Instead, we find that the photometric uncertainties are overestimated by a factor of $\approx$2 (i.e., $\sigma_j\approx0.5$) in the case of sources brighter than $m\sim15$, and underestimated by a factor of 1.2--1.3 (i.e., $\sigma_j\approx0.8$) for sources fainter than $m\sim17$ and brighter than $m\sim18.5$.

\begin{figure*}
	\includegraphics[width=0.45\linewidth]{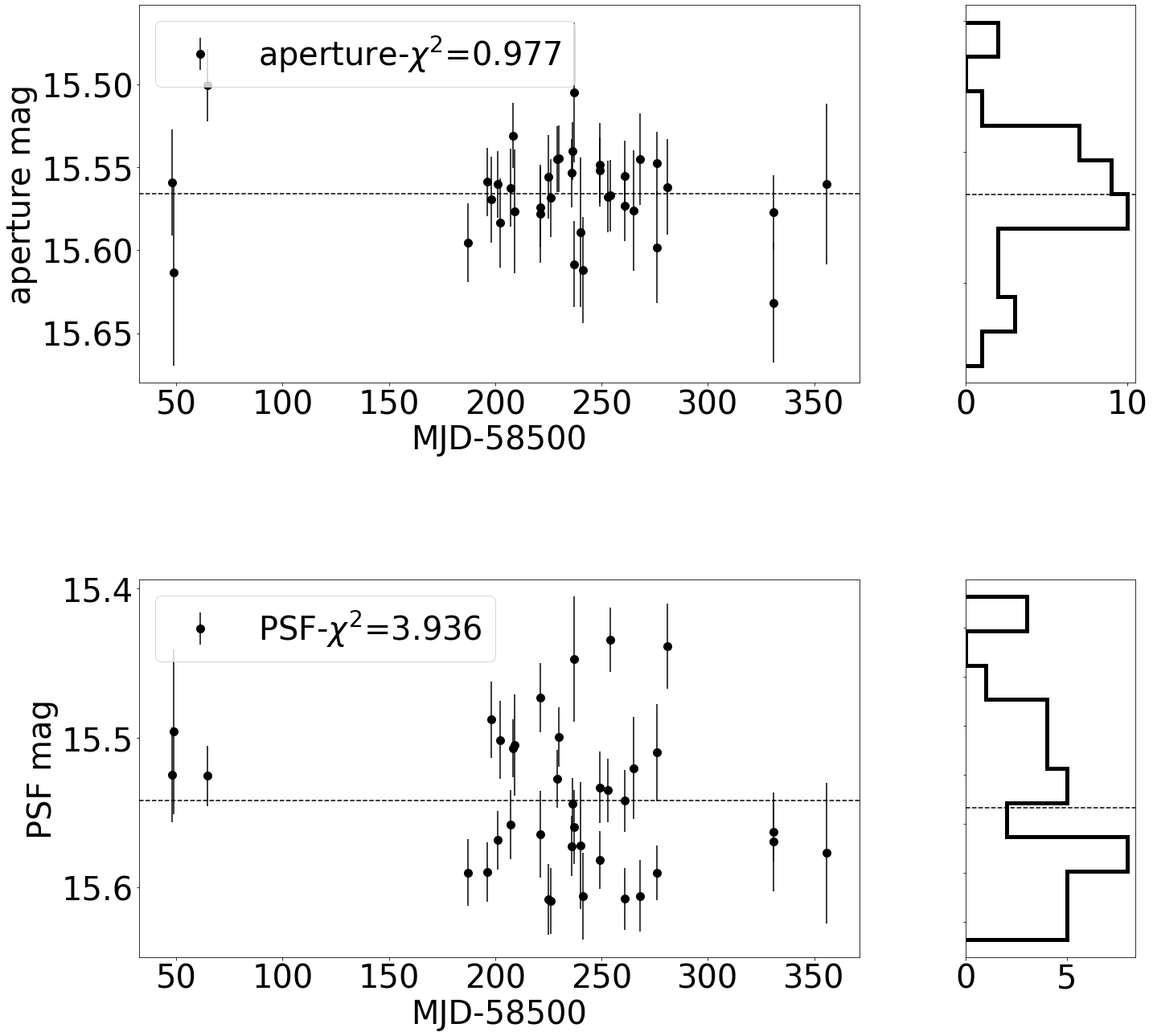}
	\includegraphics[width=0.45\linewidth]{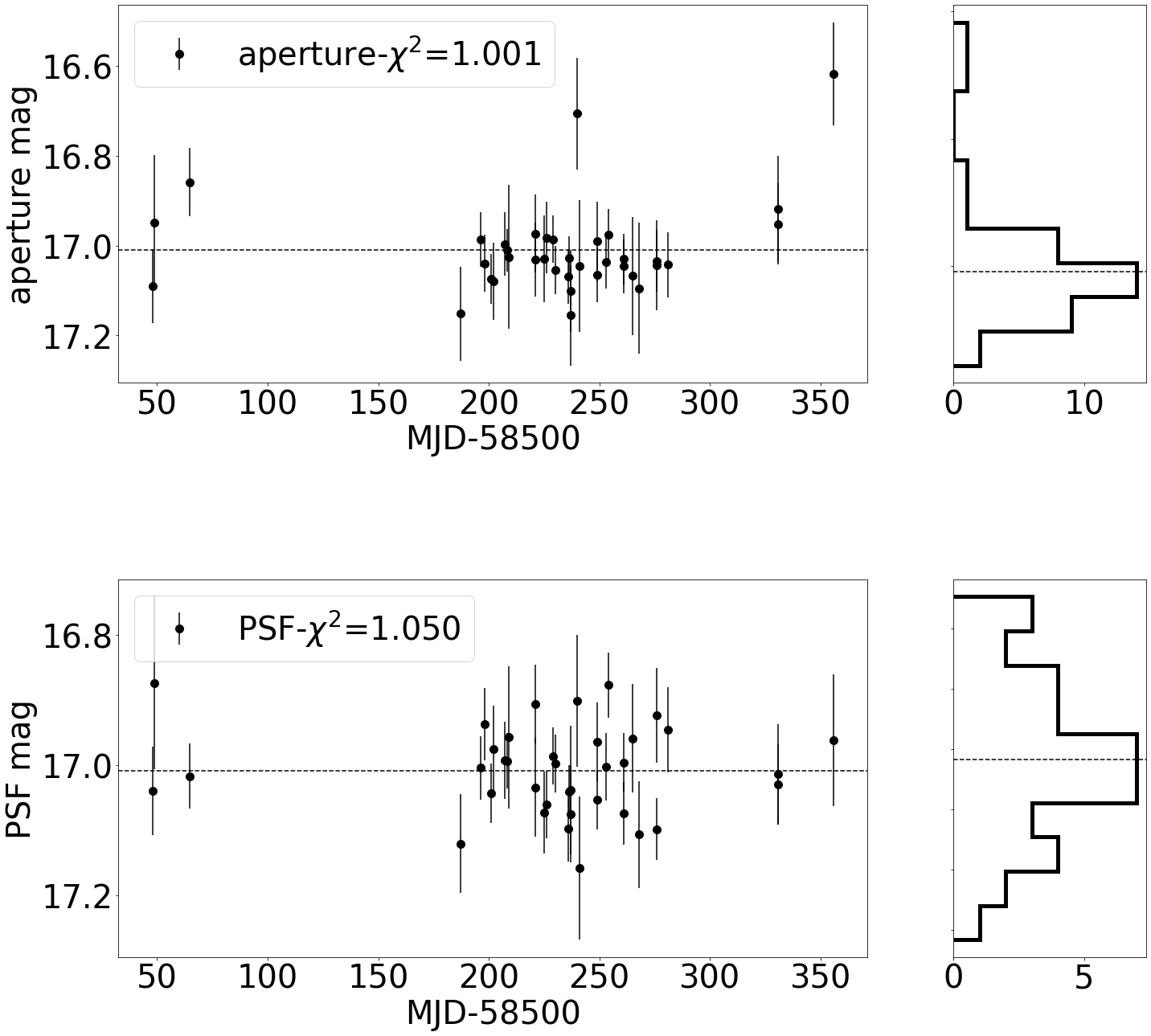}
	\centering
    \caption{Lightcurves of standard SDSS stars from aperture (upper panels) and PSF (lower panels) photometry. Aperture photometry gives reduced-$\chi^2$ closer to unity, which is what is expected for non-variable stars. Aperture photometry performs better at brighter (i.e., < 15 mag) magnitudes.}
    \label{fig:LC1}
\end{figure*}

In the middle panel of Figure \ref{fig:errorplots} the standard deviation of the magnitude difference (again calculated using Eq. \ref{sdev}, but with $\Delta m_{ij}=m_{ij}-\overline{m_{ij}}$ in place of $z_{ij}$), which we denote $\sigma_j(\Delta m)$. Finally, in the bottom panel of Figure \ref{fig:errorplots} we plot the median error of each of our sources, again averaged over all observations, $i$. From these lower two panels, we find that the standard SDSS stars brighter than $m\sim15$ have a standard deviation in $\Delta_m$ that is less than $\approx$0.015 whereas the minimum uncertainty for these magnitudes is $\approx$0.032. Again, this implies that the uncertainty estimate is too large by a factor of $\approx$2 in this bright regime.

%RLCS: the caption is not very clear to me. Is the top panel the stddev of the value z from eq 2, where the stddev is calculated as in eq 3? -DONE 

Since the aforementioned results are based on standard deviation measurements, it is important to ensure that they are not affected by artificial factors such as the number of observations (e.g., the standard deviation will only start to approximate to the size of the uncertainty after a large number of measurements). To test for this, we explore whether the trends seen in the top plot of Figure \ref{fig:errorplots} changes as a function of the number of epochs (see Figure \ref{fig:errorepochs}). However, while there is perhaps some evidence of larger uncertainties at fewer epochs, this effect is very weak and certainly not large enough to explain the trend seen in Figure \ref{fig:errorplots}.

Our results therefore support the application of correction factors to the error bars, such as those presented in \cite{Suberlak2017} for the CRTS data, prior to using the photometric uncertainties when studying source variability. In our case, we calculate the correction factor by fitting a 4\textsuperscript{th} degree polynomial to the median values of $\sigma_L$ shown in the top plot of Figure \ref{fig:errorplots}. The resulting polynomial has the following terms:
\begin{equation}
0.0074 x^4 - 0.5168 x^3 + 13.42 x^2 - 153.4 x + 651.5
\end{equation}

We then use this polynomial to correct our uncertainties. These corrected uncertainties are included in the following section where we present some examples of lightcurves measured using forced photometry on GOTO data by the LSST stack.

\subsection{Light curve analysis}

As a final approach to characterise the forced photometry results, we examine some light curves generated using the nightly catalogues generated from our forced photometry measurements. First, we compare the light curves generated using aperture photometry against those using PSF photometry. For standard SDSS stars (\citealt{Ivezic2007}) brighter than $\sim$17.5 we find that the aperture photometry light curves are better, in terms of reduced-$\chi^2$ after single flux fitting of the lightcurves,  than the light curves obtained from PSF photometry, especially at the very bright end (< 15 mag; see Fig.\ref{fig:LC1}). We find, however, that for stars fainter than $\sim$17.5 both types of photometry give similar values of reduced-$\chi^2$, with the PSF photometry perhaps giving slightly better measurements, as implied by Fig. \ref{fig:psfvsAp}.

%\begin{figure*}
%	\includegraphics[width=0.45\linewidth]{figures/lc1}
%	\includegraphics[width=0.45\linewidth]{figures/lc2}
%	\centering
%    \caption{Lightcurves of variable star from aperture and PSF photometry. The general shape of the PSF light curves follow the trend of the aperture photometry light curve. For the PSF light curves, however, the plots show an addtional intra-night scatter.}
%    \label{fig:LC2}
%\end{figure*}

\begin{figure*}
	\includegraphics[width=0.45\linewidth]{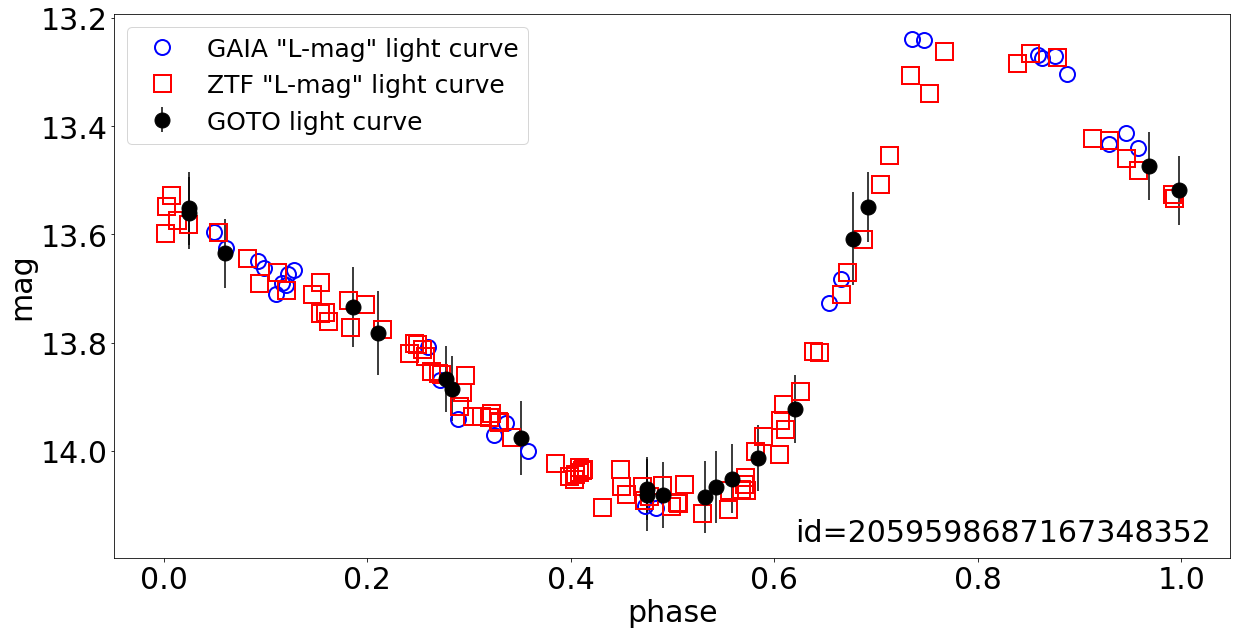}
	\includegraphics[width=0.45\linewidth]{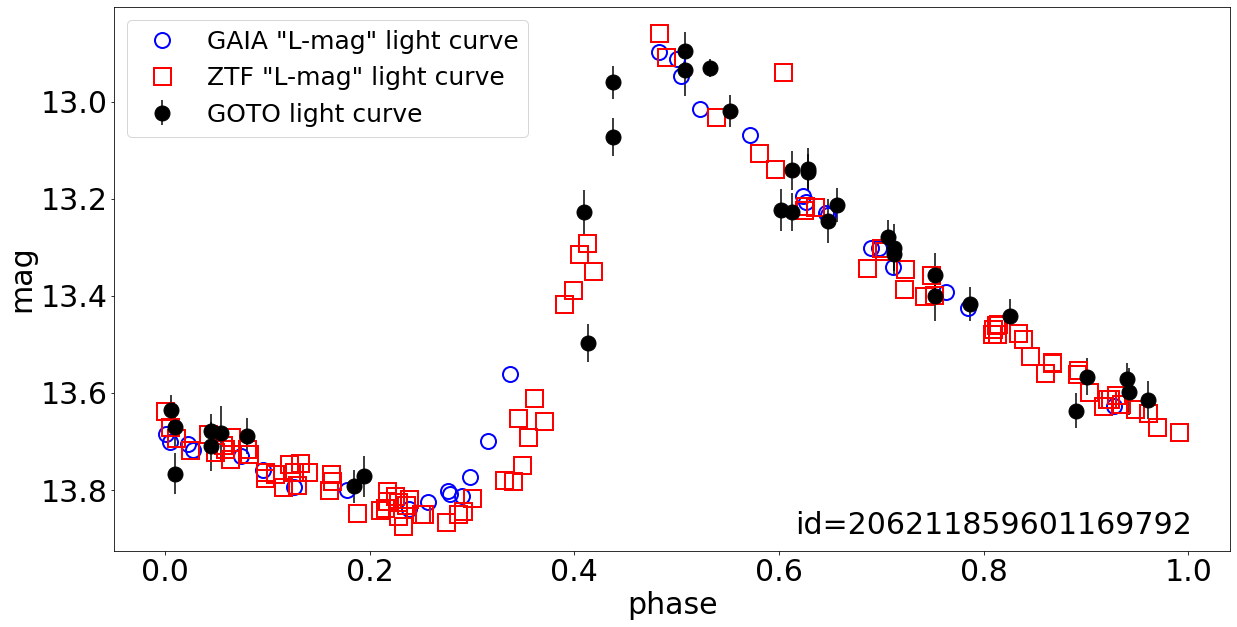}
	\includegraphics[width=0.45\linewidth]{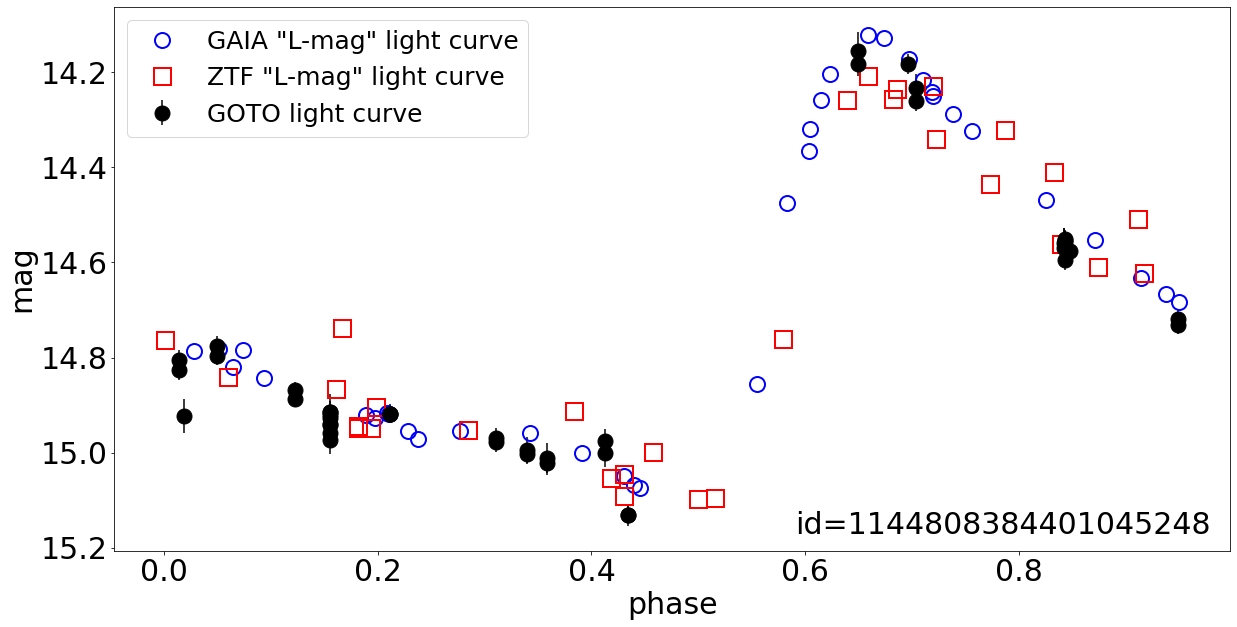}
	\includegraphics[width=0.45\linewidth]{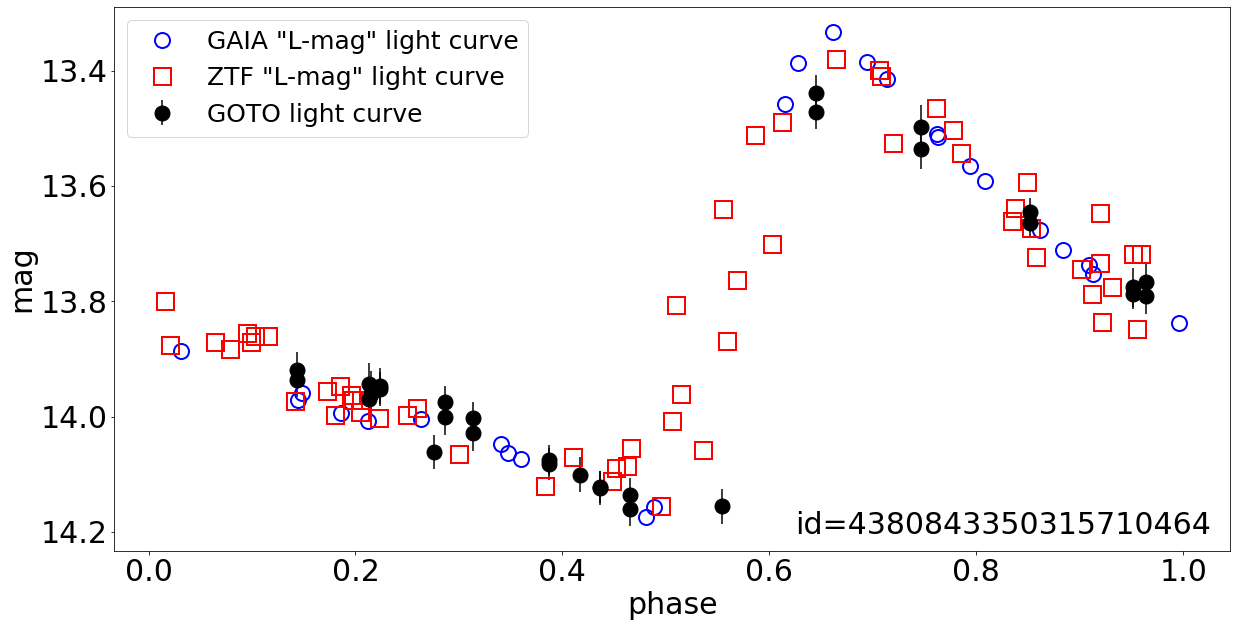}
	\centering
    \caption{Examples of GOTO L-band light curves from aperture photometry measured with the LSST stack, presented as phase plots, of periodic variable stars. Also included in these plots are ZTF g-band and GAIA G-band light curves, colour-corrected to L-band. These variable stars have period from 0.5 to 6 days and belong to two different classes (top row: Cepheid variables, bottom row: RR Lyraes, the id numbers refer to GAIA ids). GOTO light curves contain fewer data points since it has (a) been operating for a shorter period of time than the other surveys and (b) the ZTF in particular has a higher cadence than the GOTO prototype. The shapes of the GOTO, ZTF and GAIA phase plots are very similar, demonstrating that GOTO will be a valuable resource for time-domain studies of variable sources once it has its full complement of 16 UTs in both the Northern and Southern hemispheres.}
    \label{fig:gotoztf}
\end{figure*}

Finally, to characterise the performance of the forced photometry performed by the LSST stack task on GOTO data, we compare our light curves against those for the same sources obtained by the ZTF and GAIA surveys. ZTF is a wide-field survey that uses the Palomar 48 inch Schmidt telescope and with a dedicated camera of 47 deg$^2$ field of view (\citealt{Graham2019}). Its large FOV allows ZTF to scan the northern sky with a 3-day cadence in g and r bands. It also scans the visible Galactic plane every night. The survey has a median limiting g-band and r-band magnitude of $\sim$20.8 and $\sim$20.4, respectively. The ZTF survey is similar to GOTO's high-cadence survey, although GOTO's cadence with 4 UTs is lower than that of ZTF. However, when GOTO is fully deployed (with 16 UTs in both the northern and southern hemispheres), its cadence will match that of ZTF. By contrast, GAIA is a space based mission whose primary goal was to measuring accurate positions, parallaxes, and proper motions of over 10 billion stars (\citealt{Gaia16}). To achieve this goal, it observed those stars multiple times during its nominal five-year mission and, in doing so, obtained multi-epoch photometry measurements.

We select four different variable stars (two Cepheids and two RR Lyrae stars) from the General Catalogue of Variable Stars (GCVS 5.1; \citealt{Samus2017}) with periods of 0.5 to 6 days (these periods are those reported by that same catalogue). The ZTF light curves are generated using data from their second data release (DR4; \citealt{Masci19}) and only include epochs from the public survey. We present the results in figure \ref{fig:gotoztf}. The GAIA lightcurves are generated from the second GAIA data release (DR2; \citealt{Gaia18}). We have colour corrected both the ZTF g-band and and GAIA G-band photometry to create synthetic ZTF and GAIA ``L''-band photometry. The light curves in this case are given as phase plots in which repeated observations of the same part of the cycle are effectively ``folded''. We calculate the phase of a given observation by subtracting a reference starting time (the same start time is used in the case of both GOTO, ZTF, and GAIA), dividing by the period of the source, then taking the remainder fraction of the period. We note that that after applying the correction for the errors as implied by Fig. \ref{fig:errorplots}, the uncertainties generated by our modified forced photometry task are larger than those on the ZTF and GAIA light curves.

For these comparison plots for the GOTO lightcurves we have used the LSST stack aperture photometry as produced by the forced photometry task whereas ZTF uses PSF photometry. The larger number of data points on the ZTF and GAIA light curves are partly due to both having operating for a longer period than we are considering in this study (i.e., ZTF DR4 covers the period between March, 2018 and June, 2020, and GAIA covers the period between 25 July 2014 and 23 May 2016 ), and especially ZTF's higher cadence over this time. Having said that, it is clear the GOTO lightcurves track the shape of their respective ZTF and GAIA light curves very closely. As such, this comparison make us confident that GOTO will be a valuable resource for measuring variable sources, particularly when it has its full complement of UTs.

\section{Summary}

In this paper we have used the LSST stack to perform forced photometry on images obtained with the GOTO prototype. This has involved the development of the \texttt{obs\_goto} package which works as the interface between the GOTO data and the LSST stack. The \texttt{obs\_goto} package is described in more detail in  \cite{Mullaney20}, together with a description of the production of a series of reference images and subsequent reference catalogue for a large  fraction (i.e., $\sim50\%$) of the GOTO-observable sky.

In this paper we have presented the nightly processing of the data through the forced photometry task. We used the reference catalogue described \cite{Mullaney20} to obtain the positions of sources for which we measure the flux within a number of apertures. We have also performed PSF photometry for these sources. From the nightly forced photometry catalogue we were then able to generate light curves for the sources. We then assessed the quality of the measured forced photometry by comparing our results to those obtained by {\sc gotophoto} -- GOTO's own processing pipeline, whose photometry measurement is based on SExtractor -- Pan-STARRS, ZTF, and GAIA.

After comparing against colour-corrected Pan-STARRS g-band photometry, we found that our L-band photometric measurements were consistent to within 0.01 mag (rms) for brighter sources ($\sim$14 mag) to 0.2 mag (RMS) for fainter sources ($\sim$ 18 mag). We also performed internal photometric tests by assessing the consistency of repeated measurements of standard stars. This assessment showed that the typical precision for bright (i.e. $<$ 15.5), unsaturated sources is 0.02 mag. This assessment also indicated that the GOTO aperture photometry from the LSST stack is more precise than PSF photometry which has a precision of 0.04 mag for bright sources. The survey depth of a GOTO pointing ($\sim$ 19.4 mag), again measured via the repeatability of aperture photometry, is found to be slightly brighter than that reported ($\sim$19.6 mag) for the reference catalogue described in \cite{Mullaney20}. Finally, by comparing the measured uncertainties to the standard deviation of repeat-observed sources, we found that the photometric errors associated with brighter sources are overestimated by a factor of $\sim$2, whereas they are underestimated by a factor of $\sim$1.3 for sources fainter than $\sim17$. 

Our results demonstrate the feasibility of using the LSST stack to process and perform forced photometry measurements on GOTO data. In section 3.1 we have described the steps we took to adapt the LSST stack to process and mean-combine GOTO data from a single pointing and perform forced photometry on the resulting frame. This adaptation is included in our \texttt{obs\_goto} package. This highlights a particular advantage of using the LSST stack -- i.e., that the user can modify or even write tasks as part of the ``obs\_package’’ to process their own data in the way they wish, while exploiting the various modules that make up the LSST stack. Finally, the results from our various quality assurance tests demonstrate that the data obtained via the LSST stack's processing of GOTO frames can be used for the scientific analysis of light curves.

\section*{Acknowledgements}

We thank the referee for their constructive comments, which helped to clarify some key points. We also thank those members of the LSST Community Forum whose help and advice enabled us to conduct this study. The Gravitational-wave Optical Transient Observer (GOTO) project acknowledges the support of the Monash-Warwick Alliance; Warwick University; Monash University; Sheffield University; the University of Leicester; Armagh Observatory \& Planetarium; the National Astronomical Research Institute of Thailand (NARIT); the University of Turku; Portsmouth University; and the Instituto de Astrof\'{i}sica de Canarias (IAC). This paper makes use of software developed for the Large Synoptic Survey Telescope. We thank the LSST Project for making their code available as free software at http://dm.lsst.org. Based on observations obtained with the Samuel Oschin 48-inch Telescope at the Palomar Observatory as part of the Zwicky Transient Facility project. ZTF is supported by the National Science Foundation under Grant No. AST-1440341 and a collaboration including Caltech, IPAC, the Weizmann Institute for Science, the Oskar Klein Center at Stockholm University, the University of Maryland, the University of Washington, Deutsches Elektronen-Synchrotron and Humboldt University, Los Alamos National Laboratories, the TANGO Consortium of Taiwan, the University of Wisconsin at Milwaukee, and Lawrence Berkeley National Laboratories. Operations are conducted by COO, IPAC, and UW. R.P.B., M.R.K. and D.M-S. acknowledge support from the ERC under the European Union’s Horizon 2020 research and innovation programme (grant agreement No. 715051; Spiders).

%%%%%%%%%%%%%%%%%%%%%%%%%%%%%%%%%%%%%%%%%%%%%%%%%%



%\bibliographystyle{mnras}
\bibliographystyle{pasa-mnras}
\bibliography{gotolsstfp}
\end{document}